\documentclass[11pt,prd,superscriptaddress,nofootinbib]{revtex4}
\usepackage[english]{babel}
\usepackage{graphicx}
\usepackage{mathtext}
\usepackage{indentfirst}
\usepackage{epsfig,amsmath,amsfonts}

\newcommand{\beqn}{\begin{eqnarray}}
\newcommand{\eeqn}{\end{eqnarray}}

\newcommand{\bear}{\begin{array}}
\newcommand{\enar}{\end{array}}

\begin{document}

\hfill ITEP--TH--17/14

\vspace{5mm}

\centerline{\Large \bf Secularly growing loop corrections in strong electric fields}

\vspace{5mm}

\centerline{E. T. ${\rm Akhmedov}^{1),2),3)}$, N. ${\rm Astrakhantsev}^{2),3)}$ and F.K.${\rm Popov}^{2),3)}$}


\begin{center}
{\it $\phantom{1}^{1)}$ National Research University Higher School of Economics, International Laboratory of Representation Theory and Mathematical Physics}
\end{center}

\begin{center}
{\it $\phantom{1}^{2)}$ Institutskii per, 9, Moscow Institute of Physics and Technology, 141700, Dolgoprudny, Russia}
\end{center}

\begin{center}
{\it $\phantom{1}^{3)}$ B. Cheremushkinskaya, 25, Institute for Theoretical and Experimental Physics, 117218, Moscow, Russia}
\end{center}

\vspace{3mm}

\begin{center}{\bf Abstract}\end{center}
We calculate one--loop corrections to the vertexes and propagators of photons and charged particles in the strong electric field backgrounds. We use the Schwinger--Keldysh diagrammatic technique. We observe that photon's Keldysh propagator receives growing with time infrared contribution. As the result, loop corrections are not suppressed in comparison with tree--level contribution. This effect substantially changes the standard picture of the pair production. To sum up leading IR corrections from all loops we consider the infrared limit of the Dyson--Schwinger equations and reduce them to a single kinetic equation.

\vspace{5mm}

\section{Introduction}

Since the seminal paper of Schwinger \cite{Schw} pair creation by strong electric fields has been extensively studied by many authors (see e.g. \cite{Gitman1} -- \cite{Grib}). These studies were based either on tree--level calculations and/or were using Feynman diagrammatic technique. Common wisdom is that loop corrections should not bring anything substantially new to the Schwinger's pair creation picture. In fact, usually it is believed that loop contributions cannot bring anything else but the UV renormalization or corrections to the effective Lagrangian. The goal of this note is to show that this is an incorrect intuition and the tree--level picture or the picture provided by the Feynman technique is incomplete.

In condensed matter theory it is known that infrared (IR) loop corrections can become strong in non--stationary situations --- loop corrections can be comparable to the tree--level contributions (see e.g. \cite{LL} and \cite{Kamenev}). In this note we observe similar effects in scalar electrodynamics on strong electric field backgrounds. These effects, as we will see, substantially change the picture of the particle production in strong electric fields.

In particular, we show that the one--loop correction to the Keldysh propagator of the gauge field has a secularly growing contribution. This growth has nothing to do with the zero mass of the photon. It appears due to a change of the levels populations and signals that there is also photon production by the background field. This photon production happens simultaneously together with the charge pair production. (But we do not see such a growth in the propagators for the charged fields at the first loop.)

We observe that the secular growth of the one--loop contribution in constant field background starts right after we turn on interactions and lasts as long as the field is on. Such a situation leads to the so called adiabatic catastrophe \cite{PolyakovKrotov}: It means the impossibility to shift the moment after which the interactions are adiabatically turned on to the past infinity. I.e. in constant electric field background quantum field theory shows an inconsistency at the loop level. In the electric pulse background we observe the growth of the loop corrections only during the time period when the field is on. This situation in many respects is similar to the one which is seen in de Sitter space \cite{PolyakovKrotov}, \cite{Akhmedov:2008pu}, \cite{Akhmedov:2009be}, \cite{AkhmedovKEQ}, \cite{AkhmedovBurda}, \cite{Polyakov:2012uc}, \cite{AkhmedovGlobal}, \cite{AkhmedovPopovSlepukhin} (see \cite{AkhmedovReview} for the review). For the IR effects in electrodynamics in a bit different settings please consider \cite{Akhmedov:2009vh} and also \cite{Das:2012qz}, \cite{Das:2014lea}.

The fact that loop corrections to the propagators are not suppressed in comparison with tree--level contributions rises the question of the summation of the leading corrections from all loops. We address this issue in the section on kinetic equation. In that section we derive the kinetic equation for the photons on the strong background fields as the IR limit of the Dyson--Schwinger equation for the corresponding propagator.

\subsection{Setup of the problem}

We study massive scalar coupled to electromagnetic field in $(3+1)$ dimensions:

\beqn\label{action}
S = \int d^4 x\left[\frac12 \, |D_\mu \phi|^2 - \frac12 \, m^2 |\phi|^2 - \frac{1}{4} F_{\mu\nu}^2 - j^{cl}_\mu A^\mu\right].
\eeqn
Here $D_\mu = \partial_\mu + i e A_\mu$; the source $j^{cl}_\mu$ creates a background field, which is a solution of Maxwell's equations $\partial^\mu F^{cl}_{\mu \nu} = j^{cl}_{\nu}$. The corresponding gauge-potential is $A^{cl}_\mu$. We divide the full gauge potential into two pieces $A_\mu = A^{cl}_\mu + a_\mu$ --- classical and quantum parts.

Throughout this paper, we will denote the external gauge-potential $A^{cl}_\mu$ as $A_\mu$. We will study two particular types of background fields: the constant field $A_1(t) =  E t$, for which $j^{cl}_\mu = 0$, and  the pulse $A_1(t) = E T \tanh\left(\frac{t}{T}\right)$, which transforms into  $E t$, as $T \to \infty$.

In the presence of an external electromagnetic field the equation of motion for $\phi$ is  $\left(D_\mu^2 + m^2\right) \phi = 0$. The harmonic expansion of the scalar field is as follows $\phi(t,\vec{x}) = \int \frac{d^3 p}{(2\pi)^3}\left[a_{\vec{p}} \, e^{i \, \vec{p} \, \vec{x}} f_p(t) + b^+_{\vec{p}} \, e^{-i \, \vec{p} \, \vec{x}} f^*_{-p}(t)\right]$, where the time harmonics, $f_p(t)$, obey:

\beqn\label{eom_harm}
\left[\partial_t^{2\phantom{\frac12}} + \,\, \omega_p^2(t)\right]\,f_p(t) = 0, \quad {\rm and} \quad \omega_p(t) = \sqrt{m^2 + \left[\vec{p} + e \vec{A}(t)\right]^2}, \quad \vec{A}(t) = \left(A_1(t),0,0\right).
\eeqn
For those choices of $A_1(t)$ that we consider in this note one can find $f_p(t)$ exactly (see e.g. \cite{Grib}), but for our purposes we do not need this exact form. We will use the WKB approximation, which works when $\left|p_1 + e A_1(t)\right| \gg m$ \cite{PolyakovKrotov}. In this approximation the harmonic functions can be represented as:

\beqn\label{gen_harm}
f_p(t) = \left\{\begin{matrix}
\frac{A(p_\perp)}{\sqrt{2 \omega_p(t)}} e^{- i \int\limits^t_{t_{hc}} \omega_p(t') dt'} + \frac{B(p_\perp)}{\sqrt{2 \omega_p(t)}} e^{i \int\limits^t_{t_{hc}} \omega_p(t') dt'}, \,\, t<t_{hc}\\
\frac{C(p_\perp)}{\sqrt{2 \omega_p(t)}} e^{- i \int\limits^t_{t_{hc}} \omega_p(t') dt'} + \frac{D(p_\perp)}{\sqrt{2 \omega_p(t)}} e^{ i \int\limits^t_{t_{hc}} \omega_p(t') dt'}, \,\, t>t_{hc}.\\
\end{matrix}\right.
\eeqn
In the vicinity of the point $t_{hc}$, where $p_1 + e A_1(t_{hc}) = 0$, the WKB approximation breaks down.
Here $A,B,C,D$ are some complex functions of the momentum orthogonal to the background field, $\vec{p}_\perp = (p_2, p_3)$. When $A=1, \,\, B=0$ the corresponding harmonics, $f_p^{in}(t)$, and annihilation operators, $a^{in}_p$ and $b^{in}_p$, define the so called in-state: $a^{in}_p \left|in \right\rangle = b^{in}_p \left|in \right\rangle = 0$ (see e.g. \cite{Grib}). For such harmonics, however, both $C$ and $D$ are simultaneously not zero. (This can be seen if one considers (\ref{eom_harm}) as the Schr\"{o}dinger equation with the scattering potential $\omega^2(t)$.) Furthermore, when $C=1, \,\, D=0$ we have the so called out-harmonics, $f^{out}_p(t)$, and the corresponding annihilation operators, $a^{out}_p$ and $b^{out}_p$. They define the so called out-state. For these harmonics $A$ and $B$ are simultaneously not zero.

In the pulse background $\omega_p(t)$ is time independent when $t \ll T$ and $t \gg T$. Hence,
in this case the in-- and out--harmonics become just linear combinations of the ordinary plane waves. That is not the case in the constant electric field background. However, in the constant field
eq.(\ref{eom_harm}) has an extra symmetry: $p_1 \to p_1 + \alpha$ and $t \to t - \alpha/eE$. Then, $f_p(t) = f_{p_\perp}(p_1 + e E t) = f_{p_\perp}(p_{ph})$, where $p_{ph} = p_1 + e E t$. Furthermore, we have that $f^{in *}_{p_\perp}\left(p_{ph}\right) = f^{out}_{p_\perp}\left(-p_{ph}\right)$ (see e.g. \cite{Anderson:2013ila}, \cite{Anderson:2013zia}). Moreover, as the corollary of this symmetry in the constant electric field background one can construct a peculiar time-symmetric state \cite{Anderson:2013ila}, \cite{Anderson:2013zia}, for which harmonic functions obey $f^{s*}_{p_\perp}\left(p_{ph}\right) = f^{s}_{p_\perp}\left(-p_{ph}\right)$. For these harmonics all $A,B,C,D$ in (\ref{gen_harm}) are nonzero.

Because of the described above behavior of the harmonic functions the Hamiltonian of the free scalar theory cannot be diagonalized once and forever \cite{Grib}. However, in the pulse background in--harmonics diagonalize this Hamiltonian at the past infinity, while out--harmonics do the same at the future infinity. At the same time, in the constant electric background none of the choices of the harmonic functions does the diagonalization of the free Hamiltonian at the past or future infinity. That is because the background field is never switched off, while the free Hamiltonian is diagonalized only by single plane waves \cite{Grib}.
As the result, in the formulas below we do not specify the explicit form of the harmonic functions, $f_p(t)$, unless it is necessary to complete the calculation. Moreover, in such a situation we prefer to study the behavior of the correlation functions. Only the proper interpretation of this behavior allows sometimes to find a particle description (see e.g. \cite{AkhmedovReview}).

Because of the time--dependence of the free Hamiltonian, the field theory under consideration is in non-stationary situation. Hence, to calculate correlation functions one has to apply the Keldysh-Schwinger (KS) diagrammatic technique instead of the Feynman one \cite{LL}, \cite{Kamenev}. In such a formalism every particle is described by the matrix propagator, whose entries are the Keldysh propagator $D^K = \frac{1}{2} \left\langle\left\{\phi(x),\bar{\phi}(y)\right\}\right\rangle$, and the retarded and advanced propagators $D^{A,R} = \mp\theta(\mp\Delta t)\left\langle\left[\phi(x),\bar{\phi}(y)\right]\right\rangle$ (and the same for the gauge fields, with $\phi \to a_\mu$). Due to spatial homogeneity of the background fields we find it convenient to make the spatial Fourier transformation of these propagators. Then, at the tree--level they look like:

\beqn\label{treelev}
D^K_0\left(p,t_1,t_2\right) = \frac{1}{2}\left[f_p(t_1)\,f^*_p(t_2) + f_p(t_1)\,f^*_p(t_2)\right],\\
D^{R,A}_{0} (p,t_1,t_2) = \mp\theta(\mp\Delta t)\left[f_p(t_1)\,f_p^*(t_2)-f^*_p(t_1)\,f_p(t_2)\right],\notag\\
G^K_{0\,\mu\nu}(p,t_1,t_2) = - g_{\mu\nu}\frac{\cos\left[|p|(t_1-t_2)\right]}{2|p|} \quad {\rm and} \quad
G^{R,A}_{0\,\mu\nu}(p,t_1,t_2) =  \mp i g_{\mu\nu} \theta(\mp\Delta t)\frac{\sin\left[|p|(t_1-t_2)\right]}{|p|}.\notag
\eeqn
Propagators denoted as $D^{K,A,R}_0$ and $G^{K,A,R}_0$ describe complex scalar and electromagnetic fields, correspondingly. Unlike the standard textbooks situations we consider here exact harmonics, $f_p(t)$, rather than plane waves.

Apart from other advantages, the partial Fourier transformation allows to address the behavior of each $\vec{p}$--harmonic separately. Then the retarded and advanced propagators allow to specify the spectrum of the quasi--particles, while the Keldysh propagators specify the state of the theory, i.e. define which $\vec{p}$-levels are occupied \cite{Kamenev}. E.g., if the quantum average was done with the use of an arbitrary state $|\psi \rangle$ which respects spatial translational invariance, the form of the Keldysh propagators would have been:

\beqn\label{refpoint}
D^K(p,t_1,t_2) = \left[\left\langle\psi\left|a^+_{\vec{p}} a_{\vec{p}}\right|\psi\right\rangle + \frac12\right] \,  f_p(t_1) f_p^*(t_2) + \left\langle\psi\left|a_{\vec{p}} b_{-\vec{p}}\right|\psi\right\rangle \, f_p(t_1)f_{p}(t_2) + \nonumber \\ + (a \to b, \vec{p}\to - \vec{p}, \,\, h.c.), \nonumber \\
G^K_{\mu\nu}(q,t_1,t_2) =  \left[\left\langle\psi\left|\alpha^+_{\vec{q}\mu} \alpha_{\vec{q}\nu}\right|\psi\right\rangle - \frac{g_{\mu\nu}}{2}\right]\, \frac{e^{- i |q|(t_1-t_2)}}{2|q|} + \left\langle\psi\left|\alpha_{\vec{q}\mu} \alpha_{-\vec{q}\nu}\right|\psi\right\rangle \, \frac{e^{- i |q|(t_1+t_2)}}{2|q|} + h.c.,
\eeqn
while the form of the tree--level retarded and advanced propagators would not change. Such Keldysh propagators reduce to (\ref{treelev}) only if $|\psi\rangle$ is annihilated by the annihilation operators. Obviously quantum averages
$\langle \alpha^+ \alpha \rangle$, $\langle a^+ a \rangle$ and $\langle b^+ b \rangle$ define population
numbers, while $\langle \alpha \alpha \rangle$, $\langle a a \rangle$ (and their hermitian conjugates) define anomalous quantum averages. The behavior of these expectation values becomes peculiar only in non-stationary situations and only after turning on self--interactions. In fact, if there are no self--interactions or the situation is stationary, then these quantum averages just remain constant (or even zero), while the interesting situation is when they start to grow with time. The latter phenomenon cannot be seen in Feynman diagrammatic technique.

For the better understanding of the discussion below it is convenient to keep in mind that Schwinger--Keldysh technique is explicitly causal and loop expressions, which we obtain below with the use of this technique, can be understood as solutions of Cauchy problem whose initial data are given by the tree--level expressions.

\subsection{Tree--level current}

The most interesting correlation function to study in the strong background electric fields is the current. At the tree--level it looks as:

\beqn
\left\langle : J_x :\right\rangle = 2 e \, \int \frac{d^3p}{\left(2\pi\right)^3} \left(p_1 + eEt\right)\left[\left|f_p(t)\right|^2 - \frac{1}{2\, \omega_p(t)}\right].
\eeqn
Here the last term under the integral cancels UV divergent contribution to the current, if it is present.

In the constant electric field background  we obtain that $\left\langle : J_x : \right\rangle = 0$ for the time--symmetric vacuum. To see this one has to convert the integration variables $p_1 \to p_{ph}$
and to note that for this vacuum $\left|f^s_{p_\perp}(p_{ph})\right|^2$ is an even function of $p_{ph}$. Thus, the current should vanish just as the consequence of the time translation and time reversal invariance of the theory in the constant electric field (for the discussion on this issue see, e.g., \cite{Anderson:2013ila}, \cite{Anderson:2013zia}).

At the same time, for the pulse background the result is that \cite{Gavrilov:2007hq}, \cite{Akhmedov:2009vs}, \cite{PolyakovKrotov}, \cite{Anderson:2013ila}, \cite{Anderson:2013zia}:

\beqn
\left\langle : J_x : \right\rangle \propto T \, e^3 E^2 \, e^{-\frac{\pi \, m^2}{e E}}.
\eeqn
The physical meaning of this answer is easy to understand. If we have a situation with the Schwinger's constant pair production per unit four--volume --- $\Gamma \propto \left(eE\right)^2 \, e^{-\frac{\pi \, m^2}{e E}}$ --- then, the density of the charge carriers grows linearly and, hence, the current should also grow during the whole period, $T$, when the background field is on.

\section{One--loop contributions}

In this section we show that in the strong background electric fields there are correlation functions which have growing with time (IR) loop contributions. We start our consideration with two--point functions, $D^{K,A,R}(p, t_1, t_2)$ and $G^{K,A,R}(p,t_1,t_2)$, and continue with the vertices (three--point functions). For the case of the two--point functions we take the limit $\frac{t_1+t_2}{2} = t \to \infty$, when $t_1 - t_2 = {\rm const}$.

We would like to stress now that the partial Fourier transformed expressions for the loop integrals
are not straightforwardly sensitive to the UV loop divergences. To see the latter one has to transform back to the spatial coordinates or to make the full space-time Fourier transformation, if possible.

To start with, let us show that there are no large contributions to the retarded and advanced propagators in the limit under consideration. E.g. the one-loop contribution to the advanced scalar propagator is as follows\footnote{The argumentation for the other retarded and advanced propagators (both for photons and scalars) is absolutely the same.}:

\beqn\label{1loop}
D^A_1(p,t_1,t_2) = \int\limits^{t_2}_{t_1} dt_4 \int\limits^{t_4}_{t_1} dt_3 D^A_0(p,t_1,t_3) \, \Sigma^A(p,t_3,t_4) \, D^A_0(p,t_4,t_2),
\eeqn
where more or less standard explicit expressions for the self--energy $\Sigma^A(p,t_3,t_4)$ via $D^{K,A,R}_0$ and $G^{K,A,R}_0$ in several different theories can be found in \cite{LL} or \cite{Kamenev}. We do not need the explicit formula to draw our conclusions. In fact, because of the presence of the Heaviside $\theta$--function in every retarded and advanced propagator we have those limits of the $t_{3,4}$ integration which are shown in (\ref{1loop}). This fact is in the basis of the proof that $D_1^A$ and $D_1^R$ have the same advanced and retarded properties as their tree--level counterparts \cite{Kamenev}. Because of these properties $D^A_1$ cannot have growing contributions, if $t_1 - t_2$ is held fixed.

\subsection{One--loop correction to the photon's Keldysh propagator}

We start with the case of photons. In the limit under consideration the leading one--loop correction to the photon's Keldysh propagator can be written in the following form:

\beqn\label{GK}
G^K_{\mu\nu}(q,t_1,t_2) =  \left[n_{\mu\nu}(q,t) -  \frac{g_{\mu\nu}}{2}\right] \, \frac{e^{- i |q|(t_1-t_2)}}{2|q|}  + \kappa_{\mu\nu}(q,t)\frac{e^{- i |q|(t_1+t_2)}}{2|q|} + h.c.,
\eeqn
where:

\beqn\label{photondens}
n_{\mu\nu}(q,t) = e^2 \int \limits^t_{t_0}  dt_3 \int \limits^t_{t_0} dt_4 \frac{e^{-i |q| (t_3-t_4)}}{2|q|} \int \frac{d^3 k}{(2\pi)^3} \times \nonumber \\ \times \left[f_k(t_3) D_\mu f_{k+q}(t_3)^{\phantom{\frac12}} - \,\, D_\mu f_k(t_3) f_{k+q}(t_3)\right]\, \left[f^*_k(t_4) D_\nu f^*_{k+q}(t_4)^{\phantom{\frac12}} - \,\, D_\nu f^*_k(t_4) f^*_{k+q}(t_4)\right], \nonumber \\
{\rm and} \quad \kappa_{\mu\nu}(q,t) = - 2 e^2 \int \limits^t_{t_0}  dt_3 \int \limits^{t_1}_{t_0} dt_4 \frac{e^{i |q| (t_3+t_4)}}{2|q|} \int \frac{d^3 k}{(2\pi)^3} \times \nonumber \\ \times \left[f_k(t_3) D_\mu f_{k+q}(t_3)^{\phantom{\frac12}} - \,\, D_\mu f_k(t_3) f_{k+q}(t_3)\right] \, \left[f^*_k(t_4) D_\nu f^*_{k+q}(t_4)^{\phantom{\frac12}} -  \,\, D_\nu f^*_k(t_4) f^*_{k+q}(t_4)\right].
\eeqn
Here $D_\mu \, f_p(t) \equiv \left(\partial_t, i p_1 + i e\,A_1(t), i p_2, i p_3\right)\, f_p(t)$ and $t_0$ is a moment of time, after which the interactions are adiabatically turned on. In these expressions we neglect the difference between $t_{1,2}$ and $t$ in the limit under consideration. This is mathematically rigorous in the leading approximation, if $n_{\mu\nu}(q,t)$ and $\kappa_{\mu\nu}(q,t)$ have a divergence as $t\to +\infty$.

To better understand the point that we advocate below, please compare (\ref{GK}), (\ref{photondens}) to $G^K$ in (\ref{refpoint}). Taking into account that we have started with zero $n_{\mu\nu}$ in (\ref{treelev}) its growth with time due to loop effects would mean that quantum corrections generate photon production (the change of the level population). At the same time, secular growth of the anomalous quantum average $\kappa_{\mu\nu}$ would mean that the initial ground state of the field in question is substantially modified in the future infinity.

To calculate $n_{\mu\nu}$ and $\kappa_{\mu\nu}$ we start with the case of the constant electric field, when $f_p(t) = f_{p_\perp}(p_1 + e E t)$. Then we make the following change of integration variables $t' = \frac{t_3 + t_4}{2}, \tau = \frac{t_3 - t_4}{2}$. As the result, we obtain the $\tau$--integral in the range $[t_0-t, \, t-t_0]$, but its integrand is rapidly oscillating for large $\tau$. Hence, to estimate $n_{\mu\nu}$ and $\kappa_{\mu\nu}$ we can extend the upper and lower limits of the $\tau$--integration to the plus and minus infinity, correspondingly. Then the result is:

\beqn\label{IR_diverg_photon}
n_{\mu\nu}(q,t) = 2 \, e^2 \, \int\limits^t_{t_0} dt' \int \limits^\infty_{-\infty} d\tau \frac{e^{-2 i |q| \tau}}{2|q|} \int \frac{d^3 k}{(2\pi)^3} \times \nonumber \\ \times \left[f_{k}(\tau) D_\mu f_{k + q}(\tau)^{\phantom{\frac12}} - \,\, D_\mu f_{k}(\tau) f_{k + q}(\tau)\right] \, \left[f^*_{k}(-\tau) D_\nu f^*_{k + q}(-\tau)^{\phantom{\frac12}} - \,\, D_\nu f^*_{k}(-\tau) f^*_{k + q}(-\tau)\right], \nonumber \\
{\rm and} \quad \kappa_{\mu\nu}(q,t) = - 4 e^2 \int \limits^t_{t_0} \frac{e^{2 i |q| t'}}{2|q|}  dt'\int \limits^\infty_{0} d\tau \int \frac{d^3 k}{(2\pi)^3} \times \nonumber \\ \times \left[f_k(\tau) D_\mu f_{k+q}(\tau)^{\phantom{\frac12}} - \,\, D_\mu f_k(\tau) f_{k+q}(\tau)\right]\, \left[f^*_k(-\tau) D_\nu f^*_{k+q}(-\tau)^{\phantom{\frac12}} -  \,\, D_\nu f^*_k(-\tau) f^*_{k+q}(-\tau)\right].
\eeqn
Now one can see that the integrand of $\int dt'$ in the expression for $n_{\mu\nu}$ does not depend on $t'$. (From the mathematical point of view that happens due to the invariance of the harmonic functions $f_{k_\perp}(k_1 + eEt)$ under simultaneous compensating shifts of $k_1$ and $t$.) As the result, the two-point function has a large contribution as follows --- $n_{\mu\nu} = e^2 (t-t_0) A_{\mu\nu}$, where

\beqn
A_{\mu\nu} = 2 \, \int \limits^\infty_{-\infty} d\tau \frac{e^{-2 i |q| \tau}}{2|q|} \int \frac{d^3 k}{(2\pi)^3} \times \nonumber \\ \times \left[f_{k}(\tau) D_\mu f_{k + q}(\tau)^{\phantom{\frac12}} - \,\, D_\mu f_{k}(\tau) f_{k + q}(\tau)\right] \, \left[f^*_{k}(-\tau) D_\nu f^*_{k + q}(-\tau)^{\phantom{\frac12}} - \,\, D_\nu f^*_{k}(-\tau) f^*_{k + q}(-\tau)\right] \nonumber
\eeqn
This contribution is present for any choice of the exact harmonic functions $f_p(t)$ and it is straightforward to show that its coefficient, $A_{\mu\nu}$, is well defined in the sense of generalized functions: I.e. neither $\tau$ nor $k$ integrals are divergent, but for large $k$ and $q$ there are $\delta$--functional contributions to $A_{\mu\nu}$. At the same time, due to the presence of the oscillating factor inside the $dt'$ integration in the second equation in (\ref{IR_diverg_photon}), we do not have such a divergence in $\kappa_{\mu\nu}$. This just means that $\kappa_{\mu\nu}$ is suppressed in comparison with $n_{\mu\nu}$ and we have obviously chosen the correct ground state for the electromagnetic field. But the presence of the secularly growing $n_{\mu\nu}$ means that in the future infinity the theory in question is going to end up in an excited rather than in the photon's vacuum state\footnote{Let us stress here that the contribution in question has nothing to do with the vanishing photon's mass. In fact, let us add to the theory in question the Yukawa coupling of the charged scalar under consideration, $\phi$, to a massive real (neutral) scalar, $\lambda \varphi |\phi|^2$. Then the analogous to (\ref{IR_diverg_photon}) expression for $n_p$ of $\varphi$ is even simpler: Under the $d\tau$ integral one just has the product of four harmonic functions without derivatives. Then $n_p$ does have the same type of contribution. Obviously, however, its prefactor is suppressed for more massive fields.}.

The contribution in question is nothing but an IR divergence. Its regulator, $t_0$, cannot be taken to the past infinity. I.e., unlike the case of zero background electric field, one cannot take the moment of turning on self--interactions to the past infinity. In fact, when $E\to 0$ the harmonic functions are converted into plain waves, $f_p(t) \to e^{i \omega_p \, t}$. Then the factor of the divergence, $e^2 (t-t_0)$, becomes proportional to $\int d^3k \, \delta\left(|q| + \omega_k + \omega_{k-q}\right)$, after the $d\tau$ integration. The $\delta$--function here ensures the energy conservation. Hence, if $E=0$ the prefactor of the IR divergence in $n_{\mu\nu}$ is zero and one can take $t_0 \to -\infty$. At the same time, if the background field is not zero the sharp $\delta$--function gets eroded because there is no conservation of energy in  time--dependent background fields. As the result we obtain the IR divergence in question\footnote{Note that if one were using another gauge for the background field, where $A_0 = - Ex$, the loop calculation would have been similar. One should also use Schwinger--Keldysh technique and keep in mind that, if the situation is really stationary, the result of calculation reduces to the one obtained via the Feynman technique. In this gauge one will obtain the $\delta$--function establishing energy conservation. But the argument of such a function can become zero, because the invariant energy standing in the argument of this $\delta$--function is $\omega - eEx$. Hence, in this gauge one will also encounter the time divergence in question.}.

As we will see below, the factor multiplying $e^2 (t-t_0)$ is just a piece of the collision integral, which is due to the (unusual for the empty space) particle creation by the background field. We have here the simultaneous creation of one photon, $e^{iqt}$ in (\ref{IR_diverg_photon}), and two oppositely charged scalars, $f_k$ and $f_{k+q}$.

Note that even if $e^2$ is very small, after a long enough time period loop correction, $e^2 (t-t_0)$, becomes comparable to the tree--level contribution. I.e. the loop correction, $n_{\mu\nu}$, is essentially a classical quantity. That is not a very unusual phenomenon in non--stationary quantum field theory \cite{Kamenev}. These observations put forward the question of the summation of all unsuppressed loop corrections
in the limit $t-t_0 \to \infty$. We address this problem below in the section on kinetic equation.

The presence of such a divergence in $n_{\mu\nu}$ simply means that one cannot have eternal and everywhere constant electric field: although constant $E$ is a vacuum solution of the Maxwell's equations ($j^{cl}_\mu = 0$), and the tree--level current, $\langle : J_x : \rangle$, is zero, the quantum field theory is still shows an inconsistency at the loop level.

In the light of what we have just said, it is physically appropriate to consider electric pulse background $A_1(t) = E T \tanh\left(\frac{t}{T}\right)$ instead of the eternal and everywhere constant field. In the pulse background one obtains the same expressions for $n_{\mu\nu}$ and $\kappa_{\mu\nu}$ as in (\ref{photondens}) with the corresponding harmonic functions $f_p(t)$. Unfortunately, then the integrals in (\ref{photondens}) cannot be taken exactly, but we can estimate them when $T\to \infty$, $t_0 \ll - T$ and $t \gg T$. We can distinguish three regions of time integration: before ($t_0 < t_{3,4} < -T$), inside ($-T < t_{3,4} < T$) and after ($T < t_{3,4} < t$) the electric pulse. The interference terms between these regions do not bring large contributions to $n_{\mu\nu}$ and $\kappa_{\mu\nu}$.

Before and after the pulse, the harmonics are linear combinations of plain waves, $A e^{- i \omega_p t} + B e^{i \omega_p t}$, with some complex functions, $A$ and $B$, of $p_{\perp}$. After their substitution into (\ref{photondens}) and the change of the integration variables and, then, integration over $\tau = (t_3-t_4)/2$, the situation becomes similar to the case of zero background field: under the $d^3 k$ integral we obtain $\delta$--functions of the type $\delta(|q| \pm \omega_{k-q} \pm \omega_{k})$. The arguments of these $\delta$--functions never zero. Hence, the integration regions under discussion do not bring large contributions to $n_{\mu\nu}$. At the same time, to estimate the contribution to $n_{\mu\nu}$ coming from the region inside the pulse we can use the same calculation as in the constant field background, if $T\to \infty$. Therefore, in the pulse we have the following large IR contribution: $n_{\mu\nu} \sim e^2 \, T \, A_{\mu\nu}$.

\subsection{Properties of $n_{\mu\nu}$}

We will show now that $n_{\mu\nu}$ is transversal, $n_{\mu\nu}(q,t)\, q^\nu = 0$. It is obvious that $n_{\mu\nu}(q, t < t_0) \, q^\nu = 0$ because we have started with such a state where $n_{\mu\nu}(q, t) = 0$ before $t_0$. After the multiplication of (\ref{photondens}) by $q^\mu$ and a straightforward calculation, we find:

\beqn
n_{\mu\nu}(q,t) \, q^\mu = e^2 \int \limits^t_{-\infty} dt_3 \int \limits^t_{-\infty} dt_4 \int \frac{d^3 k}{(2\pi)^3} \frac{|q| \, e^{- i |q| (t_3-t_4)}}{2 |q|} \times \nonumber \\ \times \left[f_k(t_3) \partial_{t_3} f_{k-q}(t_3)^{\phantom{\frac12}} - \,\, \partial_{t_3} f_k(t_3) f_{k-q}(t_3)\right] \, \left[f^*_k(t_4) D_\nu f^*_{k-q}(t_4)^{\phantom{\frac12}} - \,\, D_\nu f^*_k(t_4) f^*_{k-q}(t_4)\right] + \notag \\ +
e^2 \int \limits^t_{-\infty} dt_3 \int \limits^t_{-\infty} dt_4 \int \frac{d^3 k}{(2\pi)^3} \frac{ e^{- i |q| (t_3-t_4)}}{2 |q|} \times \nonumber \\ \times i \partial_{t_3} \left[f_k(t_3) \partial_{t_3} f_{k-q}(t_3)^{\phantom{\frac12}} - \,\, \partial_{t_3} f_k(t_3) f_{k-q}(t_3)\right] \, \left[f^*_k(t_4) D_\nu f^*_{k-q}(t_4)^{\phantom{\frac12}} - \,\, D_\nu f^*_k(t_4) f^*_{k-q}(t_4)\right]  = \notag\\
= i e^2 \int \limits^t_{-\infty} dt_4 \int \frac{d^3 k}{(2\pi)^3} \frac{e^{- i |q| (t-t_4)}}{2 |q|} \times \nonumber \\ \times \left[f_k(t) \partial_{t} f_{k-q}(t)^{\phantom{\frac12}} - \,\, \partial_{t} f_k(t) f_{k-q}(t)\right] \, \left[f^*_k(t_4) D_\nu f^*_{k-q}(t_4)^{\phantom{\frac12}} - \,\, D_\nu f^*_k(t_4) f^*_{k-q}(t_4)\right].
\eeqn
To derive this expression we have used equations of motion for $f_p(t)$. Now we use $f_p(t) = f_{p_\perp}(p + e E t)$ and make the change of variables $t_4 = t + \tau, \,\, k \to k - e E t$. Then:

\beqn
n_{\mu\nu}(q,t) q^\mu = i e^2 \int \limits^0_{-\infty} d\tau \int \frac{d^3 k}{(2\pi)^3} \frac{e^{ i |q|  \tau_1}}{2 |q|}\times \nonumber \\ \times \left[f_k(0) \partial_{t} f_{k-q}(0)^{\phantom{\frac12}} - \,\, \partial_{t} f_k(0) f_{k-q}(0)\right]\left[f^*_k(\tau) D_\nu f^*_{k-q}(\tau)^{\phantom{\frac12}} - \,\, D_\nu f^*_k(\tau) f^*_{k-q}(\tau)\right].
\eeqn
Hence, $n_{\mu\nu} \, q^\mu$ is time independent and $n_{\mu\nu}(q,t) q^\mu = n_{\mu\nu}(q,t<t_0) q^\mu = 0$. Thus, we have that $n_{\mu\nu}(q, t) = \pi_{\mu\nu} \, n_q(t)$, where $\pi_{\mu\nu}$ is a time independent symmetric transversal tensor $|q| \pi_{0\nu} - q_i \pi_{i\nu} = 0$. Note, however, that this proof does not quite work for the case of the pulse background due to the moments of turning on and switching off the background field. At the same time, we do not know any deep physical reason why $n_{\mu\nu}$ has to be transversal in all situations. All we can state at this point is that $n_{\mu\nu}$ receives only transversal contributions from the constant field background.

Furthermore, it can be seen form (\ref{photondens}) that $n_q(t)$ is positive, because:

\beqn
n_q(t) \propto e^2 \int \frac{d^3 k}{(2\pi)^3} \frac{1}{2|q|}\left|\int \limits^t_{t_0} dt \left[f_k(t) D_\mu f_{k+q}(t)-D_\mu f_k(t) f_{k+q}(t)\right] e^{-i |q| t}\right|^2 \ge 0.
\eeqn
Finally we would like to understand if the presence of non--zero $n_{\mu\nu}$ does lead to a flux of photons or does not. To see that, we calculate the quantum average of $T_{0i}$ component of the energy momentum tensor:

\beqn
\left\langle : T_{0i} : \right\rangle \equiv \left\langle T_{0i}(E) \right\rangle - \left\langle T_{0i}(E=0) \right\rangle =
\int \frac{d^3 q}{(2\pi)^3} \frac{1}{2 \left|\vec{q}\right|} \left[q_0 q_i \, n + q_\mu^{2\phantom{\frac12}} n_{0i} - q_\mu q_i \, n_0^\mu - q_0 q^\mu \, n_{\mu \, i}\right].
\eeqn
Using that $n_{\mu\nu}\, q^\mu = 0$, $q^2_\mu = 0$ and $q^0 = \left|\vec{q}\right|$ for the photon, we find $\left\langle :T_{0i}: \right\rangle = \int \frac{d^3 q}{(2\pi)^3} \frac{n(q)}{2} q_i$. It is not hard to see that $n(-\vec{q}) = n(\vec{q})$. Hence $\left\langle : T_{0i}: \right\rangle = 0$ and the flux is seemingly vanishing. (The diagonal components of $\langle : T_{\mu\nu} : \rangle$ do not vanish.)
However, one can check that $\langle : T_{01}^{R} : \rangle$, which contains the integral $\int_0^{+\infty} dq_1$, is not zero. This is the flux in the right direction along the $x$--axis. Also $\langle : T_{01}^L : \rangle$, which contains the integral $\int_{-\infty}^0 dq_1$, is also not zero. But they compensate each other is the total expression for the flux. Hence, the flux is actually not zero but is equal in both positive and negative $x$--directions simultaneously.

\subsection{One--loop correction to the scalar Keldysh propagator}

The one--loop correction to the scalar Keldysh propagator, in the limit $t = (t_1 + t_2)/2 \to \infty$, when $t_1 - t_2 = const$, can be also expressed as:

\beqn
D^K(p,t_1,t_2) = \left[n^{+\phantom{\frac12}}_p(t) + \frac{1}{2}\right] f_p(t_1) f_p^*(t_2) + \kappa_p^+(t) f_p(t_1)f_{p}(t_2) + \left(+ \leftrightarrow -, \vec{p} \leftrightarrow - \vec{p}\right),
\eeqn
where:

\beqn\label{quantum_quantity_part}
n^{+}_p(t) = e^2 \int\limits^t_{t_0}dt_3 \int\limits^t_{t_0}dt_4\int\frac{d^3q}{(2\pi)^3} \frac{e^{-i|q|(t_3-t_4)}}{2|q|} \times \nonumber \\ \times \left\{-\left[\left(2\vec{p}_\perp + \vec{q}_\perp\right)^2 + (2p_1 + q_1 + 2 e E t_3)(2p_1 + q_1 + 2 e E t_4)\right]f_p(t_3)f^*_p(t_4)f_{p+q}(t_3)f^*_{p+q}(t_4) + \right.\notag\\\left. + \left[\dot{f}_p(t_3)f_{p+q}(t_3) - f_p(t_3) \dot{f}_{p+q}(t_3)\right]\, \left[\dot{f}_p^{*}(t_4)f_{p+q}^*(t_4) - f_p^*(t_4) \dot{f}_{p+q}^{*}(t_4)\right]\right\}, \\
{\rm and} \quad \kappa^{+}_p(t) = e^2 \int\limits^t_{t_0}dt_3\int\limits^t_{t_0}dt_4 \int\frac{d^3q}{(2\pi)^3} \frac{e^{-i|q|(t_3-t_4)}}{2|q|} \times \nonumber \\ \times \left\{-\left[\left(2\vec{p}_\perp + \vec{q}_\perp\right)^2 + (2p_1 + q_1 + 2 e E t_3)(2p_1 + q_1 + 2 e E t_4)\right]f_p(t_3)f^*_p(t_4)f_{p+q}(t_3)f^*_{p+q}(t_4) + \right.\notag\\\left. + \left[\dot{f}^{*}_p(t_3)f_{p+q}(t_3) - f_p^*(t_3) \dot{f}_{p+q}(t_3)\right] \, \left[\dot{f}_p^{*}(t_4)f_{p+q}^*(t_4) - f_p^*(t_4) \dot{f}_{p+q}^{*}(t_4)\right]\right\}, \notag
\eeqn
and similar expressions for the anti-particles --- $n^{-}$ and $\kappa^-$. Here we denote $\dot{f}(t) = df/dt$ and, to trace the existence of the divergence, we neglect the difference between $t_{1,2}$ and $t$ in the limit under consideration.

Again we start with the case of the constant electric field. Making the same transformations as in the calculation of $n_{\mu\nu}$, $t' = (t_3 + t_4)/2$ and $\tau = (t_3 - t_4)/2$, it is straightforward to show that $n^+_p(t) = n^+_{p_\perp}(p_{ph})$, where $p_{ph} = p_1 + eEt$. Then, taking $\partial/\partial p_{ph}$ derivative of the both sides of the first equation in (\ref{quantum_quantity_part}), we obtain:

\beqn\label{par_den_der}
eE \, \frac{\partial n^+_{p_\perp}(p_{ph})}{\partial p_{ph}} \equiv I[p_{ph}]= e^2 \int \limits^\infty_{-\infty} d\tau \int \frac{d^{3} q}{(2\pi)^{3}} \frac{e^{-2 i |q|\tau}}{2|q|}
 \times \nonumber \\ \times \left\{-\left[\left(2\vec{p}_\perp - \vec{q}_\perp\right)^2 + (2p_{ph} - q_1 + 2 e E \tau)(2p_{ph} - q_1 - 2 e E\tau)\right]\right.\times \nonumber \\ \times f_{p_{\perp}}\left(p_{ph} + eE\tau\right) f^*_{p_{\perp}}\left(p_{ph} - eE\tau\right) f_{p_{\perp} - q_\perp}\left(p_{ph} + q_1 + eE\tau\right) f^*_{p_{\perp}-q_\perp}\left(p_{ph} - q_1 - eE\tau\right) + \notag\\ \left[\dot{f}_{p_{\perp}}\left(p_{ph} + eE\tau\right) f_{p_{\perp}-q_\perp}\left(p_{ph} - q_1 + eE\tau\right) - f_{p_{\perp}}\left(p_{ph} + eE\tau\right) \dot{f}_{p_{\perp}-q_\perp}\left(p_{ph} - q_1 + eE\tau\right)\right] \times \nonumber \\ \times \left. \left[\dot{f}_{p_{\perp}}^{*}\left(p_{ph} - eE\tau\right) f_{p_\perp - q_\perp}^*\left(p_{ph} - q_1 - eE\tau\right) - f_{p_{\perp}}^*\left(p_{ph} - eE\tau\right) \dot{f}_{p_{\perp}-q_\perp}^{*}\left(p_{ph} - q_1 - eE\tau\right)\right]\right\},
\eeqn
where we have changed $\vec{q} \to - \vec{q}$.
Now we would like to find the largest non--oscillating contribution to $I[p_{ph}]$ in the limit $p_{ph}\to \infty$ ($t\to + \infty$). E.g., if such a contribution is constant, then in such a limit we have a linear growth in $n^+$.

Large photon momenta, $|q|$, and large $\tau$ do not give substantial contributions to $I[p_{ph}]$ due the rapid oscillating factors under the integral. Hence, in the limit $p_{ph} \to \infty$ we can expand over powers of $q_1$ and keep only linear terms. The generic form the harmonic functions in the limit $p_{ph}\to \infty$ is as follows (\ref{gen_harm}):

\beqn\label{inter}
f_{p_\perp}(p_{ph}) \approx \alpha \cdot \left(\frac{p_{ph}}{m}\right)^{i\frac{m_\perp^2}{2 e E}} \frac{\exp\left[i \frac{p_{ph}^2}{2 e E}\right]}{\sqrt{2} \left(m_\perp^2 + p_{ph}^2\right)^{\frac14}} + \beta \cdot \left(\frac{p_{ph}}{m}\right)^{-i\frac{m_\perp^2}{2 e E}}\frac{\exp\left[-i \frac{p^2_{ph}}{2 e E}\right]}{\sqrt{2}\left(m_\perp^2 + p_{ph}^2\right)^{\frac14}}.
\eeqn
Here $m_\perp^2 = \vec{p}^2_\perp + m^2$, and $\alpha$ and $\beta$ are functions of $p_\perp$.

In the limit under consideration the largest non-oscillating contributions to $I[p_{ph}]$ will come from the interference terms between two exponents of (\ref{inter}) in the products of the harmonics functions. Then, after a straightforward calculation and, keeping only the largest and homogeneous contributions as $p_{ph} \to \infty$, we get

\beqn
I[p_{ph}] \propto e^2 \int \limits^\infty_{-\infty} d\tau \int \frac{d^{3} q}{(2\pi)^{3}}\frac{1}{p_{ph}^2} \frac{e^{-2 i |q|\tau}}{2|q|}\times \nonumber \\ \times \left[\left|\alpha(p_\perp - q_\perp)\right|^2|\beta(p_\perp)|^{2\phantom{\frac12}} e^{- 2 i q_1 \tau} + |\alpha(p_\perp)|^2|\beta(p_\perp - q_\perp)|^{2\phantom{\frac12}} e^{ 2 i q_1 \tau}  \right] \propto \frac{1}{p_{ph}^2}.
\eeqn
Obviously $\int^{p_{ph}} \frac{dk_{ph}}{k_{ph}^2}$ converges as $p_{ph} \to \infty$. Hence, there are no growing with $t$ (or $p_{ph}$) contributions to $n^+$ at the first loop. Similarly we do not find secular growth in $\kappa^+$ and in $n^-$, $\kappa^-$. It is worth stressing at this point, however, that at the second loop level ($\sim e^4$) there will be secularly growing contributions to $n^\pm$ and $\kappa^\pm$. They are coming from those divergences, which are present in $n_{\mu\nu}$ at the first loop. As we will see below, these contributions are due to the decay of the produced photons into charged pairs under the influence of the background field\footnote{It is probably worth stressing here that $n^\pm$ receive finite non--zero contributions form the first loop, but there is no secular growth.}.

We continue with the pulse background. In the region $|t_{3,4}| \ll T$, where the field is on, the situation is similar to the constant electric field background. At the same time, in the region $|t_{3,4}| \gg T$ the harmonics are just linear combinations of plain waves, $e^{\pm i \omega_p t}$. Then after the integration over $\tau$ we obtain that the corresponding contribution to $n^+$ is a linear combination of the $\delta$--functions $\delta\left(|q| \pm \omega_p  \pm \omega_{p+q}\right)$. Hence, in the pulse background we also do not have a secular growth in $n^\pm$ from the first loop. However, of course there are secularly growing contributions from the higher loops, which appear due to\footnote{It is worth stressing at this point that the behavior of $n^\pm$ depends on the type of the interaction. Consider e.g. $\lambda |\phi|^4$ selfinteraction on top of the electromagnetic one. Then $n^\pm$ will have an extra contribution at the second loop --- at $\lambda^2$ order. This contribution will contain terms which after the pulse (when the field is already switched off) will be linear combinations of the $\delta$--functions as follows: $\delta(\omega_p \pm \omega_{k_1} \pm \omega_{k_2} \pm \omega_{k_3})$. That is due to interference terms between plane waves in the products of four harmonics involved. The arguments of these $\delta$--functions can vanish, when two $\omega$'s have positive sign, while the other two are coming with the negative one. Hence, in this case we will have growing with time contributions to $n^\pm$ at the $\lambda^2$ order. It will describe creation of four particles directly from the background field.} $n_{\mu\nu}$.

In view of what we have been saying above, this may sound very surprising: on the one hand, at the tree--level we see growing current, $\langle J_x\rangle$, which should be due to created particles, but, on the other hand, we do not see a linear growth in $n^\pm$ at the first loop. To resolve the apparent paradox, one should keep in mind that the correct quantity to interpret as the particle density is $n^+_p(t)\,f_p(t)f^*_p(t)$ rather than just $n^+_p(t)$ itself. But even if $n^\pm$ is zero we still obtain a non-trivial $\langle J_x \rangle$ from zero--point fluctuations $\frac12 f_p(t)f^*_p(t)$. Note that for photons these two expressions for the particle density do coincide, because their harmonics are just plane waves.

\subsection{One--loop corrections to the vertex}

In this subsection we discuss the three-point functions in the Schwinger's, $\pm$, parametrization of the propagator matrix \cite{LL}, \cite{Kamenev}. Let us consider, e.g., $G^{---}_\mu(x_1,x_2,x_3)$. We are interested in the limit when $t_i-t_j = {\rm const}$ and $\frac{t_1+t_2+t_3}{3} \to \infty$. We make the Fourier transformation over the spatial coordinates. To understand the result of the one loop correction to the vertex, we start our consideration with the tree--level expression for $G^{---}_\mu(t,p,q)$:

\beqn\label{vertex_tree_level}
G^{---}_{\rm tree\,\mu}(t,p,q) =  \frac{e}{2|q|} {\rm Im}\left[e^{- i |q| t} f_p(t)f_{p+q}(t) \int \limits^t_{-\infty} d\tau e^{ i |q| \tau} \left(f^*_p(\tau)\overleftrightarrow{D}_\mu(\tau) f^*_{p+q}(\tau)\right) \right],
\eeqn
where $p,q$ are momenta of one of the charged scalars and of the photon, correspondingly. Here we neglect the difference between $t_{1,2,3}$ and $t = \frac{t_1+t_2+t_3}{3}$ in the limit under consideration.

At the same time the typical term appearing in the one--loop correction to $G^{---}_\mu(t,p,q)$ is

\beqn\label{vertex_loop_level}
\Delta G^{---}_{\rm loop\,\mu}(t,p,q) \propto \frac{e^3}{2|q|} \, e^{\pm i |q| t} \, \bar{f}_p(t)\bar{f}_{p+q}(t) \, \int_{-\infty}^t dt_1 e^{\mp i|q| t_1} \int \frac{d^3 k}{(2\pi)^3} \left(\bar{f}_{p-k}(t_1)\overleftrightarrow{D}_\mu(t_1)\bar{f}_{p-k+q}(t_1)\right) \times \notag\\ \times \left[\int_{-\infty}^t dt_2 \int_{-\infty}^t dt_3 \frac{e^{ \pm i|k|(t_2 -t_3)}}{2|k|} \left(\bar{f}_p(t_2) \overleftrightarrow{D}_\nu(t_2) \bar{f}_{p-k}(t_2)\right) \left(\bar{f}_{p-k+q}(t_3) \overleftrightarrow{D}^\nu(t_3) \bar{f}_{p+q}(t_3) \right)\right]
\eeqn
where $\bar{f}_p(t)$ can stand for either $f_p(t)$ or $f^*_p(t)$ depending on which particular term we consider.
After the same calculations as in the previous section, one can see, that the expression in the square brackets is convergent and is proportional to $\int (A + B\tau) d\tau e^{i |k|(1-\cos\theta) \tau} = A \delta\left[|k|(1-\cos\theta)\right] +  B \delta'[|k|(1-\cos\theta)]$, where $\theta$ is the angle between $\vec{k}$ and the background field $E$, while $A$ and $B$ are some finite expressions. Now one can see that the one--loop correction to the three--point function $G_\mu^{---}(t,p,q)$ does not receive any large IR contributions. Similar arguments are valid for other types of vertex functions, because they contain the same type of terms as in (\ref{vertex_loop_level}).

\section{Summation of the leading IR corrections from all loops (kinetic equation)}

Although $e^2$ is small, the product $e^2(t-t_0)$ (or $e^2 T$) becomes large as $t-t_0 \to \infty$ ($T\to\infty$). Hence, higher loops are not suppressed in comparison with the tree-level contribution to the photon's Keldysh propagator. To understand the physics in the strong electric fields one has to sum up leading IR contributions from all loops. We would like to perform the summation of those terms which are powers of $e^2(t-t_0)$ and to drop terms, which are suppressed by higher powers of $e$. In order to do that, we have to solve the system of Dyson-Schwinger equations for the exact propagators, $D^{K,R,A}$ and $G^{K,R,A}$, and for the vertexes in the IR limit ($t-t_0 \to \infty$ or $T\to +\infty$).

Taking into account that all vertexes, retarded, advanced propagators and also the Keldysh propagator for the scalars receive subleading corrections, we can put them to their tree--level values in the system of Dyson--Schwinger equations. Then, if we are interested only in the leading corrections, this system reduces to the single equation for the exact Keldysh propagator of the gauge field:

\beqn\label{DS_eq}
G^K_{\mu\nu}(p,t_1,t_2) = G^K_{0\,\mu\nu}(p,t_1,t_2) + \sum_{s_1, s_2, s_3 = \pm} e^2\int \limits^\infty_{-\infty} dt_3\int \limits^\infty_{-\infty} dt_4 \int \frac{d^3 k}{(2\pi)^3} \times \notag \\  \times G^{s_1 s_2}_{0\,\mu\rho}(p,t_1,t_3) D^{s_2 s_3}(p-k,t_3,t_4) \overleftrightarrow{D}_\rho(t_3) \overleftrightarrow{D}_\sigma(t_4)D^{s_2 s_3}(k,t_3,t_4) G^{s_3 s_1}_{\sigma\nu}(p,t_4,t_2),
\eeqn
where $G^{K}_{\mu\nu}$ is the exact propagator, while $G^{K}_{0\mu\nu}$ is its initial (tree--level) value.
To solve this equation we express $G^{\pm\pm}$ and $D^{\pm\pm}$ via $G^{A,R,K}$ and $D^{A,R,K}$ \cite{LL}, \cite{Kamenev} and then use the ansatz (\ref{GK}) for the exact $G^K$. We put all the rest of propagators to their tree--level values and set $\kappa_{\mu\nu} = 0$ in (\ref{GK}), because it does not receive large corrections at the tree--level. At the same time for $G^K_0$ we also use (\ref{GK}) with $\kappa^0_{\mu\nu} = 0$ and $n_{\mu\nu}^0 \neq 0$.

We would like to pick out the largest IR contribution from the integral on the RHS of (\ref{DS_eq}). The calculation is just a straightforward generalization of the one performed in the previous section.  Finally, one can convert the integral DS equation into the integrodifferential form, i.e., into the form of the kinetic equation.
After the extraction of the largest contribution to the RHS of (\ref{DS_eq}), this is done as follows:

\beqn\label{kin_eq}
\frac{n_{\mu\nu} - n_{\mu\nu}^0}{t-t_0} \to \frac{\partial n_{\mu\nu}(q,t)}{\partial t} = - \Gamma_{1\,\mu}^\rho (q) \, \left[-g_{\rho\nu}^{\phantom{\frac12}} + \,\, n_{\rho\nu}(q,t)\right] + \Gamma_{2\,\mu}^\rho (q) \, n_{\rho\nu}(q,t), \notag\\
{\rm where}\quad \Gamma_{1\,\mu \nu}(q) = e^2 \int \frac{d^3 k}{(2\pi)^3} \int\limits^\infty_{-\infty} d\tau \frac{e^{-2 i |q| \tau}}{|q|} \times \nonumber \\ \times \left[f_k(\tau) D_\mu f_{k-q}(\tau)^{\phantom{\frac12}} - \,\, D_\mu f_k(\tau) f_{k-q}(\tau)\right] \, \left[f^*_k(-\tau) D_\nu f^*_{k-q}(-\tau)^{\phantom{\frac12}} - \,\, D_\nu f^*_k(-\tau) f^*_{k-q}(-\tau)\right] \notag \\
{\rm and} \quad \Gamma_{2\,\mu \rho}(q)=  e^2 \int \frac{d^3 k}{(2\pi)^3} \int\limits^\infty_{-\infty} d\tau \frac{e^{-2 i |q| \tau}}{|q|}\times \nonumber \\ \times \left[f^*_k(\tau) D_\mu f^*_{k-q}(\tau)^{\phantom{\frac12}} - \,\, D_\mu f^*_k(\tau) f^*_{k-q}(\tau)\right]\, \left[f_k(-\tau) D_\rho f_{k-q}(-\tau)^{\phantom{\frac12}} - \,\, D_\rho f_k(-\tau) f_{k-q}(-\tau)\right].
\eeqn
One can check that $\Gamma_{1,2 \, \mu\nu}(q)$ are transversal $\Gamma_{1,2 \, \mu\nu}(q) q^\nu = 0$. E.g., for $\Gamma_1$ this was done in the previous sections, because if we put $n_{\mu\nu} = 0$ on the RHS of (\ref{kin_eq}) we reproduce the one--loop result. Thus, $\Gamma_{1,2}$ can be represented as $\Gamma_{1,2 \,\mu\nu}(q) = \pi_{\mu\nu} \, \Gamma_{1,2}(q)$, where $\pi_{\mu\nu}$ is the above defined symmetric transversal tensor. Also from the one-loop result for $n_q(t)$ one can see that $\Gamma_1 \ge 0$. Similarly one can show that $\Gamma_2$ is also grater than zero. Finally it is straightforward to show that $\Gamma_{1,2}(q) = \Gamma_{1,2}(-q)$.

Thus, taking the trace of $(\ref{kin_eq})$, we get the following kinetic equation for $n_q(t)$:

\beqn\label{simpl_kin_eq}
\frac{\partial n_q(t)}{\partial t} = \Gamma_1(q) \left[1 + n_q(t)\right] - \Gamma_2(q)\, n_q(t).
\eeqn
The physical meaning of the RHS of this equation is very simple. The first term describes the photon production
by the background field. The second term describes the decay of the produced photons into charged pairs.
We do not obtain on the RHS the terms describing other types of the processes because they are suppressed by higher powers of $e^2$, as we have seen above in the one--loop calculation for the scalar Keldysh propagator.

\section{Discussion and acknowledgements}

Thus, we see that vacuum of the electrodynamics can behave as a laser, if one switches on strong background field as the laser pumping. I.e. background field produces photons along with charged pairs. In fact, if one considers the limit $e\to 0$, $E\to \infty$ so that $eE = const$, then the Schwinger's probability remains finite. But the current $\langle J_x \rangle$ vanishes. To have non--zero current one has to keep $e$ finite. Then the current is not zero at the $e^1$ order --- at the tree--level. At the same time, at the one--loop, $e^2$, order we obtain that photons density grows with time. As the result we have the photon production simultaneously with charged pairs. Note that the radiation of photons by the created charged particles is suppressed by higher powers of $e$.

To understand completely the behavior of the photon density in the strong field background, one has to solve the
kinetic equation (\ref{simpl_kin_eq}). We are not yet in a position to do that, because we can not analytically estimate $\Gamma_{1,2}$. Note that if $\Gamma_2 < \Gamma_1$ then the kinetic equation does not have a stationary solution. Otherwise there is a stationary solution for $n_q(t)$. That would mean that eternally and everywhere constant electric field is allowed by quantum field theory (after summation of all leading contributions).

We would like to acknowledge discussions with S.Apenko, V.Losyakov, A.Semenov and P.Arseev. Especially we would like to thank S.Gutz for valuable discussions. The work of ETA was done under the partial support by the grant RFBR 14-01-90405 and by the grant of the Dynasty foundation. The work of FKP is done under the partial support of the RFBR grants 14-02-31768, 14-02-31446 and by the grant of the Dynasty foundation.

\end{document}